\documentclass[amsmath,amssymb,aps,pra,reprint,notitlepage,twocolumn,showpacs,10pt,superscriptaddress,floatfix,nofootinbib]{revtex4-1}

\usepackage{graphicx}
\usepackage{bm}
\usepackage{hyperref}
\usepackage{siunitx}
\usepackage{braket}
\usepackage{microtype}
\usepackage{yfonts}
\usepackage{color}

\DeclareMathOperator{\sinc}{sinc}

\begin{document}
\title{Interference-assisted resonant detection of axions}
\author{H. B. Tran Tan}
\affiliation{School of Physics, University of New South Wales, Sydney, New South Wales 2052, Australia}
\author{V. V. Flambaum}
\affiliation{School of Physics, University of New South Wales, Sydney, New South Wales 2052, Australia}
\affiliation{Helmholtz Institute Mainz, Johannes Gutenberg University, 55099 Mainz, Germany}
%\author{C. Rizzo}
%\affiliation{Laboratoire National des Champs Magnetiques Intenses, UPR 3228, CNRS-UPS-UJF-INSA, F-31400 Toulouse Cedex, France}
\author{I. B. Samsonov}
\affiliation{School of Physics, University of New South Wales, Sydney, New South Wales 2052, Australia}
\affiliation{Bogoliubov Laboratory of Theoretical Physics, JINR, Dubna, Moscow region 141980, Russia}
\author{Y. V. Stadnik}
\affiliation{School of Physics, University of New South Wales, Sydney, New South Wales 2052, Australia}
\affiliation{Helmholtz Institute Mainz, Johannes Gutenberg University, 55099 Mainz, Germany}
\author{D. Budker}
\affiliation{Helmholtz Institute Mainz, Johannes Gutenberg University, 55099 Mainz, Germany}
\affiliation{Department of Physics, University of California, Berkeley, California 94720-7300, USA}
\date{\today}
\begin{abstract}
\begin{center}
\textbf{Abstract}\\
\end{center}
Detection schemes for the quantum chromodynamics axions and other axion-like particles in light-shining-through-a-wall (LSW) experiments are based on the conversion of these particles into photons in a magnetic field. An alternative scheme may involve the detection via a resonant atomic or molecular transition induced by resonant axion absorption. The signal obtained in this process is second order in the axion-electron interaction constant but may become first order if we allow interference between the axion-induced transition amplitude and the transition amplitude induced by the electromagnetic radiation that produces the axions.
\end{abstract}

\maketitle

\section{Introduction}

The axion is a light pseudoscalar particle proposed by Peccei and Quinn in 1977 to resolve the strong CP (charge and parity) problem in quantum chromodynamics (QCD) \cite{PhysRevLett.38.1440,Peccei2008,PhysRevLett.40.223,PhysRevLett.40.279}. Since then, the axion and other feebly interacting pseudoscalar particles with similar properties (axion-like particles or ALPs) have been identified as possible candidates to explain the observed Dark Matter (DM). Despite numerous theoretical speculations, there is still no definitive experimental evidence for the existence of these particles. The reason for this lack of evidence is two-fold. The first difficulty arises from the fact that the coupling constants of the interactions of axions \footnote{In this paper, we will not distinguish between the axion and other axion-like particles. The word `axion' will refer to both.} with Standard-Model (SM) particles, although not known precisely, are constrained to be small. As a result, any attempt to detect axions must seek to enhance the effects of the interactions and render them observable. This task is formidable. 
The second difficulty is that the axion's mass is poorly constrained so experiments that search for axions must cover a large range of frequencies. In recent years, significant efforts, both theoretical and experimental, have been made to investigate the possible parameter spaces in mass and coupling strengths.

Traditional searches for axions are based mainly on the interaction between axions and photons in the presence of a magnetic field. In such a situation, the mixing of the axion and photon states is possible and the two types of particles can be interconverted with one another \cite{PhysRevLett.51.1415,PhysRevD.32.2988}. Helioscope experiments including Sumico \cite{MORIYAMA1998147, INOUE200218, INOUE200893, OHTA201273, 1475-7516-2013-07-013}, CAST \cite{ZIOUTAS1999480, 1475-7516-2009-02-008,Collaboration2017}, SOLAX \cite{PhysRevLett.81.5068, GATTONE199959}, COSME \cite{MORALES2002325}, DAMA \cite{BERNABEI20016,doi:10.1142/S0217751X06030874,Bernabei2010}, CDMS \cite{PhysRevLett.103.141802, PhysRevLett.106.131302, PhysRevLett.111.251301, PhysRevLett.112.241302} and IAXO \cite{1475-7516-2011-06-013,1748-0221-9-05-T05002} convert solar axions into photons for detection. Haloscope experiments such as ADMX \citep{PhysRevLett.104.041301,PhysRevLett.105.171801,PhysRevD.84.121302}, HAYSTAC \cite{ALKENANY201711,PhysRevLett.118.061302} and ORGAN \cite{MCALLISTER201767} convert cosmic axions into photons in microwave cavities and detect these photons resonantly with SQUIDs. A notable feature of haloscope experiments is the long scanning time: since the energy of the incoming axions is not known, these experiments have to sweep a large frequency range to find a resonance. Light-shining-through-a-wall (LSW) experiments including ALPS \cite{Ehret:2007cm,EHRET200983,EHRET2010149}, OSQAR \cite{PhysRevD.78.092003,Pugnat2014,PhysRevD.92.092002} and GammeV \cite{PhysRevLett.100.080402} involve converting photons into axions, passing the resulting beam through a wall, which blocks all the photons but not the axions, and then converting the transmitted axions back into photons for detection on the other side of the wall. These LSW experiments do not involve frequency scanning since the energy of the axion is known from energy conservation. However, since the axion signals in these experiments scale to the fourth power in the axion-photon coupling constant (instead of second power as in helio- and haloscope experiments), the sensitivity is greatly compromised. Finally, experiments like PVLAS \cite{CANTATORE1991418, PhysRevD.77.032006,DellaValle2016}, Q \& A \cite{ALES} and BMV \cite{Battesti2008,Jackel:2010zza} search for optical birefringence (difference in optical refractive indices for different polarizations) and dichroism (difference in absorption of light of different polarizations) due to interconversion with axions \cite{MAIANI1986359}.

Recently, various new schemes for axion detection have been proposed. These include searching for axions by converting them into magnons in a ferromagnet \cite{BARBIERI1989357,BARBIERI2017135}, by looking for parity- and time-reversal-invariance violating effects (due to couplings of axions to SM particles) such as oscillating electric dipole moments \cite{PhysRevD.84.055013,PhysRevX.4.021030,PhysRevD.88.035023,PhysRevD.89.043522,PhysRevX.7.041034}, by using dielectric haloscopes (improved sensitivity compared to traditional haloscopes) \cite{1475-7516-2013-04-016,PhysRevD.88.115002,1475-7516-2017-01-061,PhysRevLett.118.091801}, by using nuclear magnetic resonance to search for axion-mediated CP-violating forces \cite{PhysRevLett.113.161801}, by resonantly detecting the oscillating magnetic flux sourced by the axions entering a static magnetic field \cite{PhysRevLett.117.141801}, by using electron spin resonance in a magnetized media to detect the oscillating effective magnetic field caused by axions \cite{1742-6596-718-4-042051}, by using superconductors \cite{PhysRevD.94.015019} and semiconductors \cite{PhysRevD.95.023013}, by using an $LC$ circuit (Dark Matter radio) \cite{PhysRevLett.112.131301,PhysRevD.92.075012,7750582}, by using Josephson junctions \cite{PhysRevLett.111.231801,BECK20156}, by using axion-induced resonant molecular transitions \cite{Arvanitaki:2017nhi}, by using laser-spectroscopy techniques to probe axion-induced atomic and molecular tranistions \cite{Braggio:2016saq}, by looking for axion-induced topological Casimir effect \cite{PhysRevD.96.015013} and by using a photon field (instead of a magnetic field) to trigger axion-to-photon decay then detecting the product photon with Raman scattering \cite{Yoshimura:2017ghb}. In this paper, we propose a new experimental scheme which is based on atomic or molecular transitions due to the absorption of axions.

The idea of using atomic transitions to produce and detect axions dates back to a 1988 paper by Zioutas and Semertzidis \cite{ZIOUTAS198894}. The authors proposed using an M1 transition to \textit{produce} axions which would then be detected in a microwave cavity. In 2014, Sikivie extended this idea and proposed using the axion-induced M1 transitions to detect galactic-halo axions \cite{PhysRevLett.113.201301}. In Sikivie's scheme, atoms in the ground state $0$ \textit{absorb} axions and go to an excited state $i$. A laser is then used to further excite the atoms to the state $f$. This laser must be tuned so that it can only cause the $i \rightarrow f$ transition. Photons emitted when the atoms in the state $f$ decay are then detected. Following Sikivie's paper, an experimental realization that uses Zeeman states in molecular oxygen at a temperature of 280 mK was proposed \cite{1367-2630-17-11-113025}. In this particular proposal, the transition frequency is scanned by applying a strong magnetic field and the detection is done via resonant multiphoton-ionization (REMPI) spectroscopy.

A general feature of the processes considered in the atom- or molecule-based proposals above is the quadratic dependence of the detection rate on the axion-electron coupling constant. Since this constant is small, the detection rate is minuscule. In this paper, we present two schemes which allow for the interference between the axion- and photon-induced atomic transition amplitudes. This interference term is linear in the axion-electron coupling constant and so we expect a significant enhancement in the signal. %As will be shown later, 
Our proposed experiment has comparable sensitivity to existing helioscope experiments \cite{1475-7516-2013-05-010}.%is sensitive to the product of the axion-photon and axion-electron coupling constants of the order of $\SI{e-17}{\per\square\giga\electronvolt}$. This is in agreement with the current observational constraints on this product, which is from $10^{-18}$ to $\SI{e-15}{\per\square\giga\electronvolt}$, depending on the axion's mass \cite{1475-7516-2013-05-010}.

\section{Experimental schemes}

\begin{figure}[tb]
\centering
\def\svgwidth{0.7\columnwidth} 
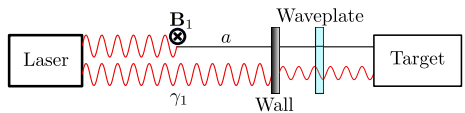
\caption{Set-up of the axion-photon interference experiment: Some of the $\gamma_1$ photons get converted into axions $a$ in the magnetic field $\mathbf{B}_1$. The resulting axion-photon beam is passed through a thin wall which suppresses the photon amplitude. A waveplate is inserted into the path of the beam to compensate for the phase difference between the axion and the photon-induced transition amplitudes.}\label{Setup Magnetic field}
\end{figure}

Our proposed experimental set-up to detect axions is shown in Fig. \ref{Setup Magnetic field}.
The photons produced by a high-power monochromatic laser (laser 1) pass through a strong magnetic field $\mathbf{B}_1$ where some photons get converted into axions. The axion-photon beam is then passed through a thin wall which suppresses the photon amplitude while leaving the axion amplitude intact. This suppression is necessary for keeping the axion-induced effect from being completely overwhelmed by its photon-induced counterpart. The axion-photon beam then hits a target and causes atomic transitions therein. 

These transitions can then be detected by, for example, using the method suggested in \cite{PhysRevLett.113.201301}. That is, another finely tuned laser is used to further excite the already excited atoms to some final state; the photons emitted when atoms in this final state decay are detected (these photons are not shown in Fig. \eqref{Setup Magnetic field}). By comparing the detected signals when the magnetic field $\mathbf{B}_1$ changes sign, one can detect the axion-induced transition amplitude.

There are different possible choices for the atoms in the target, corresponding to different transitions that one may be interested in. In this paper, we present two such schemes, wherein the axion-induced transitions are of $M0$ or $M1$ type.

\subsection{Transitions of \texorpdfstring{$M0$}~\ type}
\begin{figure}[tb]
\centering
\def\svgwidth{0.7\columnwidth} 
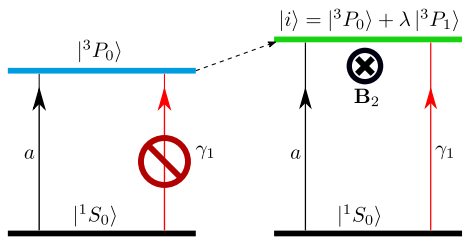
\caption{Energy levels in the target atoms in the $M0$ case. Normally, the transition $\ket{{}^1S_0} \rightarrow \ket{{}^3P_0}$ due to absorption of a photon is forbidden. If a magnetic field is applied to the target, the state $\ket{{}^3P_0}$ becomes the state $\ket{i}$ which is an admixture of $\ket{{}^3P_0}$ and $\ket{{}^3P_1}$ and the transition due to photon absorption becomes weakly allowed. The absorption of axions and photons causes a resonant transition of the target atoms from the ground state $\ket{{}^1S_0}$ to the excited state $\ket{i}$. Interference occurs between the axion- and photon-induced transition amplitudes.}\label{Level Scheme M0}
\end{figure}

The diagram for the atomic transitions in this scheme is shown in Fig. \eqref{Level Scheme M0}. The axion-induced transitions in this scheme are of $M0$ type, i.e., they happen between states of different parities and zero total angular momenta. This transition might be realized if the atoms are chosen to have two valence electrons, with the ground state $0$ being $ns^2\,{}^1S_0$ and the first excited state being $nsnp\,{}^3P_0$ (such atoms include Mg, Ca, Sr, Ba and Hg). In such atoms, the single-photon transition from the ground state ${}^1S_0$ to the excited state ${}^3P_0$ is forbidden. On the other hand, since an axion can carry the quantum numbers (angular momentum and parity) $0^-,1^+,2^-,$... \cite{PhysRevLett.40.223,ZIOUTAS198894}, this ${}^1S_0 \rightarrow {}^3P_0$ transition can be induced by absorption of axions (corresponding to axion angular momentum and parity $0^-$). 

If one applies a weak magnetic field $ \mathbf{B}_2 = B_2 \hat{\mathbf{z}}$ to the target atoms, the upper state becomes an admixture of the states ${}^3P_0$ and ${}^3P_{1,J_z=0}$:
\begin{equation} \label{admixture}
\ket{i} = \ket{{}^3P_0} 
-\frac{\bra{{}^3P_1}\boldsymbol{\mu} \cdot \mathbf{B}_2 \ket{{}^3P_0}}{E_{{}^3P_0} - E_{{}^3P_1}} \ket{{}^3P_1}\,,
\end{equation}
where $\boldsymbol{\mu} = \boldsymbol{\mu}_1+\boldsymbol{\mu}_2$ is the sum of the magnetic moments of the two valence electrons. In the presence of the applied field, there will be a one-photon transition from the ground state to the state $i$ due to the coupling between ${}^1S_0$ and ${}^3P_1$. One may choose the energy $\omega_{\gamma}$ of the photons from laser 1 (this energy equals that of the axions because the axions are produced from the same light field) so that it matches the energy difference $E_i - E_{{}^1S_0}$. In other words, the transition $0 \rightarrow i$ should be resonant. 

The total amplitude for the transition $0 \rightarrow i$ is the sum of the amplitudes due to axion absorption and photon absorption. The interference between these amplitudes allows us to detect axions. We note that for interference to occur, the photon and axion signals should have the same phase. However, as discussed in the next section, the photon absorption amplitude differs from the axion absorption amplitude by a factor of $i$ which corresponds to a phase shift of a quarter of a wavelength. To compensate for this shift, the photons should be passed through a $\lambda/4$-waveplate as shown in Fig. \ref{Setup Magnetic field}. The phase compensation should also include photon phase shift due to the thin wall. In a realistic experiment, a phase-compensating waveplate is necessary.

We point out that the $J=0 \rightarrow J'=0$ (axion) transition also exists in noble gas atoms such as Xe, Ne, Kr and Ar. This transition may happen between the ground state $np^6 \, {}^1S_0$ and the excited state $np^5\,{}^2P_{1/2}\left(n+1\right)s\left[1/2\right]_0$. The calculations for these noble gases are similar to those for the metals. We will present only the results of numerical estimates for these gases.

\subsection{Transitions of \texorpdfstring{$M1$}~\ type}
\begin{figure}
\centering
\def\svgwidth{0.7\columnwidth} 
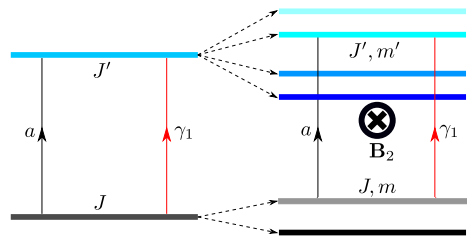
\caption{Energy levels in the target atoms in the M1 case. The absorption of axions and photons causes resonant transitions between two states with the same parity and total angular momenta differing by $0$ or $\pm 1$. Without an external field, the axion-photon interference term in the total transition probability is zero. If a magnetic field $\mathbf{B}_2$ is applied and the laser is tuned so that it induces transition between levels of specific total angular momentum projections, the axion-photon interference term can be nonzero.}\label{Level Scheme M1}
\end{figure}

A diagram for the atomic transitions in this scheme is shown in Fig. \eqref{Level Scheme M1}. As mentioned above, the axion can carry angular momentum and parity $1^+$. Thus it can induce an atomic transition of $M1$ type (total angular momentum changes by $0$ or $\pm 1$, parity does not change). Such a transition can happen between levels that have the same quantum numbers except for their total angular momenta, e.g., between the fine-structure components of the ground state.

Since a level with total angular momentum $J \neq 0$ is degenerate (with degeneracy $2J+1$ corresponding to $2J+1$ possible values of the total angular momentum projection quantum number $m$), if the target atoms are not polarized, i.e., have no definite initial projection $m$ and final projection $m'$, one needs to average the square of the total amplitude over $m$ and sum over $m'$ to get the transition probability. It is easy to see that the axion-photon interference term in this total probability vanishes. 

Indeed, as will be shown in Sect. \ref{CalculationM1}, the axion $M1$ transition amplitude is specified by the axion propagation unit vector $\mathbf{\hat k}_a$ while the photon $M1$ amplitude is proportional to the vector $\mathbf{\hat b}_{\gamma}=\mathbf{\hat k}_{\gamma}
\times \boldsymbol{\epsilon}$, where $\boldsymbol{\epsilon}$ is the photon polarization vector and $\mathbf{\hat k}_{\gamma}$ is the photon propagation unit vector (which is the same as $\mathbf{\hat k}_{a}$). For non-polarized atoms, due to spherical symmetry, the interference
term must be proportional to $\mathbf{\hat b}_{\gamma}\cdot \mathbf{\hat k}_{\gamma} $ which is zero since these two vectors are orthogonal by construction. Of course, this can be verified by straightforward calculations.

Thus, for the interference to be nonzero, it is necessary to break this spherical symmetry. This can be achieved by applying to the target a magnetic field $\mathbf{B}_2$, which defines a quantization axis, which we call the $z$-axis. The full spherical symmetry is now reduced to the rotational symmetry around this axis. This reduction of the symmetry gives a nonzero interference term.

In terms of energy levels, the field $\mathbf{B}_2$ lifts the degeneracy of the initial (total angular momentum $J$) and final (total angular momentum $J'$) levels. Sublevels with the same $J$ ($J'$) but different $m$ ($m'$) are now distinct (with energy separations of the order of $\mu_0 B_2$ where $\mu_0$ is the Bohr magneton). If one then chooses the laser frequency so that only transitions between sublevels of specific projections are induced, the axion-photon interference term should not be averaged and summed over the initial and final projections and hence no longer vanishes. We stress that the interference between the axion- and photon-induced $M1$ transition amplitudes always occurs; however, its effect averages out over the (degenerate) sublevels in non-polarized atoms. To reveal the effect of the interference, one needs to resolve these sublevels.

\section{Calculations}

\subsection{Photon-to-axion conversion probability}
The interaction between axions and photons is described by the Lagrangian density
\begin{equation}
H_{a\gamma \gamma} = -\frac{g_{a\gamma \gamma}}{4}aF_{\mu \nu} \tilde{F}^{\mu \nu}\,,
\end{equation} 
where $g_{a\gamma \gamma}$ is the axion-photon coupling constant, $a$ is the axion field, $F_{\mu \nu}$ and $\tilde{F}^{\mu \nu}$ are the electromagnetic field tensor and its dual.
This interaction is responsible for the interconversion between photons and axions in a magnetic field $\mathbf{B}_1$, with the conversion probability given in the natural units $\hbar = c =1$ by \cite{PhysRevLett.59.759,PhysRevD.39.2089}
\begin{equation}
\mathcal{P}=\frac{\omega}{4k_a}\left(g_{a\gamma \gamma}B_1 l_0\right)^2 F^2\left(q\right)\,,
\end{equation}
where $\omega$ is the photon energy (equal to the axion energy), $k_a$ is the axion's momentum, $l_0$ is the spatial extent of the magnetic field in the direction of the axion-photon beam direction, $q=\omega - k_a$ is the momentum transferred from the photon field to the axion field and $F\left(q\right) = \int{dx e^{-iqx} \frac{B\left(x\right)}{B_1 l}}$ is a form factor. For a homogeneous magnetic field $\mathbf{B}_2$, this formula becomes
\begin{equation}\label{conversion probability}
\mathcal{P} = \frac{\left(g_{a\gamma \gamma} B_1 l_0\right)^2}{4}\sinc^2\left({\frac{M^2l_0}{4\omega}}\right)\,.
\end{equation}
where $M^2 = m_a^2 + 2\omega\left(n-1\right)$, $m_a^2$ is the axion mass and $n$ is the photon refractive index.
We will not need these formulae explicitly in the calculations below. The only quantity we need for a numerical estimate is $\mathcal{P}$. 

According to the projected result of the ALPS II experiment \cite{1748-0221-8-09-T09001}, which features $l_0=\SI{100}{\meter}$ and $B_1=\SI{5.3}{\tesla}$, the upper limit on the axion-photon coupling constant $g_{\gamma a}$ ($\SI{2e-11}{\per\giga\electronvolt}$) corresponds to the photon-to-axion conversion probability $\mathcal{P} \sim 10^{-7}$. We will use these values in our estimates below.

Note that although not shown here, it is clear that the amplitude for the photon-to-axion conversion process is linear in $B_1$ (this must be true since the conversion probability is quadratic in $B_1$) and this amplitude will change its sign when $B_1$ does. We will exploit this observation to detect axions, as explained in the sections below.

\subsection{Atomic transition - the \texorpdfstring{$M0$}~\ case}
\subsubsection{Photon absorption amplitude }
The value of the coefficient $\bra{{}^3P_1,J_z=0}\boldsymbol{\mu} \cdot \mathbf{B}_2 \ket{{}^3P_0}$ in Eq.\ \eqref{admixture} is $\sqrt{2/3}\mu_0B_2$. The energy difference $E_{{}^3P_0} - E_{{}^3P_1}$ in Eq.\ \eqref{admixture} can be conveniently written as $\Delta_0 - \Delta_1$ where $\Delta_0 = E_{{}^3P_0} - E_{{}^1S_0}$ and $\Delta_1 = E_{{}^3P_1} - E_{{}^1S_0}$. The values of these energy differences in the atoms of interest are given in \cite{PhysRevA.64.012508,NIST_ASD}.

The amplitude for an atomic transition $A\rightarrow B$ induced by absorption of a photon is given by \cite{Berestetsky:1982aq,gottfried2013quantum,sobelman2012atomic}
\begin{equation}\label{General photon amplitude}
M^{\gamma} = i\sqrt{2\pi n_{\gamma}\omega_{\gamma}}e^{-i\omega_{\gamma}t}M^{\gamma}_{BA}\,,
\end{equation}
where $\omega_{\gamma}$ is the photon energy, $n_{\gamma}$ is the photon number density in the beam and $M^{\gamma}_{BA}$ is the transition matrix element. The photon energy $\omega_{\gamma}$ needs to match the difference of the energy levels $E_{i} - E_{{}^1S_0}$, corresponding to the resonant transition. In a weak magnetic field $\mathbf{B}_2$, the energy of the state $i$ is close to that of $^3P_0$ and we can take $\omega_{\gamma} \approx \Delta_0$.

In the case of the transition $0 \rightarrow i$, the leading contribution to the matrix element $M^{\gamma}_{BA}$ comes from the electric dipole (E1) term
\begin{equation}\label{Photon matrix element}
M^{\gamma}_{BA} \approx e\bra{i}\boldsymbol{\epsilon}\cdot\mathbf{r}\ket{0}=\frac{\sqrt{2}\mu_0B_2}{3\left(\Delta_0-\Delta_1\right)}\epsilon_{1z}D_S^P \,,
\end{equation} 
where $e$ is the electron charge, $\boldsymbol{\epsilon}=(\epsilon_{x},\epsilon_{y},\epsilon_{z})$ is the photon polarization vector and $D^P_S=\left<{}^3P_1\|e\mathbf{r}\|{}^1S_0\right>$ is the reduced E1 matrix element. The value of $D_S^P $ can either be calculated numerically or determined from experiment. These values for the atoms of interest are presented in \cite{PhysRevA.64.012508,NIST_ASD}.

Substituting Eq. \eqref{Photon matrix element} into Eq. \eqref{General photon amplitude}, we obtain
\begin{equation} \label{one-photon amplitude}
M^{\gamma} \approx \frac{2i}{3}\sqrt{\pi n_{\gamma}\omega_{\gamma}}e^{-i\omega_{\gamma}t}\frac{\mu_0B_2}{\Delta_0-\Delta_1}\epsilon_{z}D_S^P \,.
\end{equation}

\subsubsection{Axion absorption amplitude}

The interaction between the axion field $a$ and the electron field $\psi$ is described by the Lagrangian density
\begin{equation}\label{interaction H}
H_\textrm{int} = -\frac{g_{aee}}{2m_e} \partial_{\mu} a \bar{\psi}\gamma^5 \gamma^{\mu} \psi\,,
\end{equation}
where $g_{aee}$ is the axion-electron coupling constant. This coupling constant can be written as $g_{aee}=C_em_e/f_a$ where $f_a$ is the axion decay constant and $C_e$ is a model-dependent parameter. Currently, there are two main models for the QCD axion: the KSVZ model \cite{PhysRevLett.43.103,SHIFMAN1980493} and the DFSZ model \cite{DINE1981199,zhitnitsky1980ar}. At tree-level, $C_e=1$ for the KSVZ axion and $C_e=0$ for the DFSZ axion \cite{SREDNICKI1985689} (nonzero value $C_e \sim \alpha/2\pi \sim 10^{-3}$ appears in the latter case due to radiative corrections). Generic ALPs can have any $C_e$ value.

The amplitude for the atomic transition $A \rightarrow B$ induced by the absorption of an axion can be calculated using a method similar to that used in the case of photon absorption. We find that
\begin{equation}\label{M-axion}
M^{a}=-\sqrt{\frac{n_a\omega_a }{2}}{{e}^{-i\left({{\omega }_{a}}t+\phi_a\right)}}{{M}^a_{BA}}\,,
\end{equation}
where $\omega_a$ is the axion energy (which is equal to the energy of the $\gamma_1$ photons, $\omega_a = \omega_{\gamma}$), $\phi_a$ is the axion phase (which differs from the photon phase by the phase of the field $\mathbf{B}_1$), $n_a$ is the axion number density in the beam and the matrix element $M_{BA}$ can be derived from the interaction Hamiltonian \eqref{interaction H}. In the relativistic limit ($\omega_a \gg m_a$), the leading-order terms of $M_{BA}$ are \cite{PhysRevLett.113.201301,PhysRevD.78.115012} 
\begin{equation}\label{M_BA_exact}
\begin{aligned}
 M^a_{BA} &\approx -\frac{ig_{aee}}{2m_e}\bra{B}\mathbf{\hat k}_a\cdot \boldsymbol{\sigma}\ket{A} \\
%+\frac{iC_e\omega_a }{2f_a} \bra{B}\frac{\mathbf{p}\cdot \boldsymbol{\sigma }}{m_e}-  i\omega_a\left({\mathbf{\hat k}}_{a}\cdot \boldsymbol{\sigma }\right)\left({\mathbf{\hat k}}_{a}\cdot \mathbf{r}\right)\ket{A}\\ 
 %=-\frac{iC_e\omega_a }{2f_a}\bra{B}\mathbf{\hat k}_a\cdot \boldsymbol{\sigma}\ket{A} \\
& -\frac{g_{aee}\omega_a }{2m_e} \bra{B}\mathbf{r}\cdot \boldsymbol{\sigma }- \left({\mathbf{\hat k}}_{a}\cdot \boldsymbol{\sigma } \right) \left( {\mathbf{\hat k}}_{a}\cdot \mathbf{r} \right) \ket{A} \,,
\end{aligned}
\end{equation}
where $\mathbf{\hat k}_a = \mathbf{k}_a/\omega_a$ is axion propagation unit vector, $\mathbf{r}$ is the electron's position vector and $\boldsymbol{\sigma}$ is the electron's spin. The form of the matrix element can be qualitatively understood if we note that the relevant quantities in the problem are the electron's momentum $\mathbf{p}$ (which can be replaced by $\mathbf{r}$ by using the identity $\mathbf{p} = im_e\left[H_0,\mathbf{r}\right]$ where $H_0$ in the unperturbed electronic Hamiltonian), the electron's spin $\boldsymbol{\sigma}$ and the axion's momentum $\mathbf{k}_a$. The interaction Hamiltonian, being a pseudoscalar, must therefore be built from the scalar products of these vectors. 

It can be shown that the first term in Eq.\ \eqref{M_BA_exact} is of $M1$ type whereas the second term is of $M0$ type. For the transition  $0 \xrightarrow{a} i$, only the $M0$ term contributes to the axion amplitude. Also, for this transition, it suffices to take into account only the first term in Eq.\ \eqref{admixture} which corresponds to the state $^3P_0$. In this case,
the matrix element \eqref{M_BA_exact} can be calculated as
\begin{equation}\label{MBA}
{{M}^a_{BA}}=\frac{\sqrt{2}g_{aee}\omega_{a}R}{3m_e} \,,
\end{equation}
where $R= \int{f\left(p_{1/2}\right) f \left(s_{1/2}\right)r^3dr}$. Here, $f\left(s_{1/2}\right)$ and $f\left(p_{1/2}\right)$ are the radial parts of the upper component of the $s_{1/2}$ and $p_{1/2}$ spinor wavefunctions, respectively. The coefficient $R$ may be expressed via the reduced electric dipole matrix elements $D_S^P = \left<{}^3P_1\|e\mathbf{r}\|{}^1S_0\right>$ and $\tilde{D}_S^P = \left<{}^1P_1\|e\mathbf{r}\|{}^1S_0\right>$ as
$R=D_S^P/e + \frac1{\sqrt2}\tilde D_S^P/e$ (the $p$ electron in $nsnp\,{}^3P_1$ and $nsnp\,{}^1P_1$ states is in a linear combination of states $p_{3/2}$ and $p_{1/2}$; for $R$, we need only $p_{1/2}$ so a linear combination of $D_S^P$ and $\tilde{D}_S^P$ is necessary to eliminate $p_{3/2}$). The values of $D_S^P$ and $\tilde{D}_S^P$ for Mg, Ca and Sr are presented in \cite{PhysRevA.64.012508}. The explicit value of $R$ is given in Table \ref{Table 1}.

Substituting Eq. \eqref{MBA} into Eq. \eqref{M-axion}, we find the resulting expression for the axion absorption amplitude
\begin{equation}\label{axion amplitude}
M^{a}=-\frac{g_{aee}n_{a}^{1/2}{{\omega_{a}}^{3/2}}R}{3m_e}{{e}^{-i\left(\omega_{\gamma} t + \phi_a\right)}}\,.
\end{equation}

In contrast to the photon amplitude \eqref{one-photon amplitude}, the axion amplitude \eqref{axion amplitude} does not have the factor of $i$. This means that the phases of these amplitudes differ by $\pi/2$ in addition to $\phi_a$. To have the possibility of observing the interference between the photon and axion amplitudes, their phases need to be matched. %The phase of the photon signal may be compensated, e.g. by using a $\lambda/4$-waveplate (see Fig. \ref{Setup Magnetic field}).

\subsubsection{Axion signal and signal-to-noise ratio}%Probability of the \texorpdfstring{$\ket{s^2} \rightarrow \ket{i}$}~\ transition

Suppose that the source laser (laser 1) produces photons continuously with a rate of $N$ photons per unit time. Passing through the magnetic field, $\mathcal{P}N$ of them get converted into axions where $\mathcal{P}$ is the photon-axion conversion probability given by formula \eqref{conversion probability}. Since $\mathcal{P} \ll 1$, the number of remaining photons after conversion is approximately $N$. Denoting by $\mathcal{T}$ the photon-transmission coefficient of the wall, the number of incident photons per unit time is $\mathcal{T}N$. Thus, the photon number density in Eq.\ \eqref{one-photon amplitude} is $n_{\gamma} \propto \mathcal{T}N$ and the axion number density $n_a$ in Eq.\ \eqref{axion amplitude} is $n_a \propto \mathcal{P}N$. %Also, if the spatial extent of the target along the direction of the laser is $l$ then the total number of axions and photons present in the target at any time are $\mathcal{P}Nl$ and $\mathcal{T}Nl$, respectively.

The total amplitude for the $0 \rightarrow i$ transition is the sum of those given by Eqs.\ \eqref{one-photon amplitude} and \eqref{axion amplitude}. Squaring this sum and discarding the term which is second order in the axion-electron coupling constant, we find the total transition probability
\begin{equation} \label{P1}
\mathcal{P}\propto\left|M^{\gamma}\right|^2 + 2\operatorname{Re}\left(i\overline{M^a}M^{\gamma}\right)\,,
\end{equation} 
(the bar denotes complex conjugation). The factor $i$ comes from the inserted phase-compensating waveplate.

Multiplying this probability by the probability of transition from $i$ to $f$, one finds the detection probability. Multiplying the detection probability by the number of atoms in the target and the detection coefficient, which includes the probability of decay from the state $f$ and the sensitivity of the detector, one gets the number of observed excited atoms, which we denote by $S$. We assume that these stages of detection, associated with counting the atoms excited to state $i$, have close to 100\% efficiency.

We observe that as the magnetic field $\mathbf{B}_1$ changes sign, the phase $\phi_a$ changes by $\pi$. This corresponds to the interference term $2\operatorname{Re}\left(i\overline{M^a}M^{\gamma}\right)$ flipping sign. The same thing happens if $\mathbf{B}_2$ change its sign. Hence, the relative difference in the number of excited atoms when $\mathbf{B}_1$ or $\mathbf{B}_2$ changes sign is (here, $\Delta S$ denotes the change in $S$ as $\mathbf{B}_1$ or $\mathbf{B}_2$ flips sign)
\begin{equation}\label{Signal 1}
\begin{aligned}
\eta_{M0} &= \frac{\Delta S}{S} = \left| \frac{4\operatorname{Re}\left(i \overline{{{M}^{a}}}{{M}^{\gamma }} \right)}{\overline{{{M}^{\gamma }}}{{M}^{\gamma }}} \right|\\
&\approx \frac{2g_{aee}}{\sqrt{\pi}}\frac{\Delta_0(\Delta_0-\Delta_1)R}{m_e \mu_0 B_2 \epsilon_{z} D_S^P }\sqrt{\frac{\mathcal{P}}{\mathcal{T}}}\,.
\end{aligned}
\end{equation}
We will call this quantity $\eta$ the axion signal. Note that $\eta$ is first order in the axion-electron coupling constant.

Since the axion-electron coupling constant is small, the axion signal $\eta$ is weak. For this signal to be detectable, the contribution of the interference term in Eq. \eqref{P1} to the number of excited atoms $S$ should exceed the noise in this number. In other words, the signal-to-noise-ratio (SNR) between $\Delta S$ (the contribution of the interference term) and the noise of $S$ should be greater than unity.

Neglecting the contribution from axions, the number of excited atoms equals the number of photons absorbed by the target after the time $t$ of the experiment. This number is given by
\begin{equation}\label{S}
S \approx \mathcal{T}Nt l/l_a \,,
\end{equation}
where $l_a$ is the photon absorption length and $l$ is the length of the target. Since the fluctuation in the number of excited atoms $S$ is $\sqrt{S}$, we find
\begin{equation}\label{SNR1}
\begin{aligned}
{\rm SNR}_{M0} &= \frac{\Delta S}{\sqrt{S}} \\
&= \frac{2g_{aee}}{\sqrt{\pi}}\frac{\Delta_0(\Delta_0-\Delta_1)R}{m_e \mu_0 B_2 \epsilon_{z} D_S^P }\sqrt{\mathcal{P}Nt l/l_a} \,.
\end{aligned}
\end{equation}
Note that Eqs.\ \eqref{S} and \eqref{SNR1} are applicable when $l \ll l_a$, when the Beer-Lambert law for absorption probability $P_{\text{abs}} = 1 - \exp\left(-l/l_a\right)$ reduces to  $P_{\text{abs}} \approx l/l_a$. 

Recall that the absorption length $l_a$ is expressed in terms of the atom density in the target $n$ and the photon resonant absorption cross section $\sigma$ as \cite{Berestetsky:1982aq,gottfried2013quantum}
\begin{equation}\label{absorption length}
l_a = \frac{1}{n\sigma} \,, \qquad \sigma =  \frac{4\pi}{\Delta_0^2} \frac{\Gamma_i}{\Gamma_{\text{tot}}} \,,
\end{equation}
where $\Gamma_i = \frac{8}{81}\left(\frac{\mu_0 B_2}{\Delta_0-\Delta_1}\right)^2 \Delta_1^3\left|D_S^P\right|^2$ is the rate of the $\ket{0}\rightarrow\ket{i}$ transition and $\Gamma_{\text{tot}}$ is its total width, given by
\begin{equation}\label{Widths}
 \Gamma_{\text{tot}} \approx \left\{ \begin{array}{ll}
    \Gamma_{\text{Dop}} = 2v_0\Delta_0/\sqrt{\pi}  & \mbox{(dv)}    \,,\\
    \Gamma_{\text{col}} = 2v_0n\sigma_{\text{col}} & \mbox{(lq)}  \,,\\
\end{array} \right. 
\end{equation}
where `dv' means dilute vapor and `lq' means liquid.
Here $\Gamma_{\text{Dop}}$ is the Doppler width, while $\Gamma_{\text{col}}$ is the collisional width; $v_0 = \sqrt{2k_BT/m}$ is the most probable thermal speed of the target atoms ($k_B$ is Boltzmann's constant, $T$ is the temperature of the target and $m$ is the atomic mass) and $\sigma_{\text{col}}$ %\approx 4\pi r_{\rm atom}^2$ 
is the collisional cross section of the target atoms%($r_{\rm atom}$ is the atomic radius)
. Note that the cross section $\sigma$ in Eq.\ \eqref{absorption length} is less than the natural cross section $4\pi/\Delta_0^2$ because, due to the Doppler and/or collisional broadening, only a small fraction of the target atoms are in resonance with the laser light at any given time. The first line in \eqref{Widths} applies when the Doppler width is much larger than both the natural and the collisional widths. This condition is usually satisfied if the target medium is in a low density vapor form. In the case where the target medium is in a liquid form, the collisional width is much larger than both the Doppler and natural widths and the second line in \eqref{Widths} applies.

Substituting Eqs.\ \eqref{absorption length} and \eqref{Widths} into \eqref{SNR1}, we find
\begin{equation} \label{SNR2}
\text{SNR}_{M0} \approx 
\left\{\begin{array}{lcr}
\frac{8\pi^{1/4}g_{aee}\Delta_1^{3/2} R}{9\epsilon_{z}m_e }
\sqrt{\frac{\mathcal{P}Nlt n}{v_0\Delta_0}} &\mbox{(dv)} \,,\\
\frac{8g_{aee}\Delta_1^{3/2}R}{9\epsilon_{z}m_e}\sqrt{\frac{\mathcal{P}Nlt}{v_0\sigma_{\text{col}}}}&\mbox{(lq)} \,.
\end{array}\right. 
\end{equation}
Note that the SNR \eqref{SNR2} is independent of the transmission coefficient $\mathcal{T}$ and the magnetic field $B_2$ but is proportional to the square root of the total number of photons $Nt$ and the target size $l$. Hence, to gain a better SNR, one needs a sufficiently powerful laser and a sufficiently large target to absorb as many photons as possible. As we show in the next section, these conditions may be satisfied for noble gases in their liquid or compressed gas form but are hardly achievable for metal vapors.

Note also that the first of Eqs.\ \eqref{SNR2} has a dependence on the atom density $n$ but the second does not. This change of behavior happens at the critical value of the atom density, e.g., that of dense gases, such that the collisional and Doppler widths are comparable. As a result, as we increase the atom density, the SNR increases then saturates. After this point, we gain no further enhancement by having denser targets.

%In a totally different case where $l$ is sufficiently larger than or at least comparable to the absorption length $l_a$, $l\sim l_a$, so that $P_{\text{abs}} = 1 - \exp\left(-l/l_a\right) \approx 1$ and $N_{\text{abs}} \approx \mathcal{T}Nt$, the SNR \eqref{SNR1} reaches its maximum 
%\begin{equation}\label{SNR3}
%\begin{aligned}
%{\rm SNR_{\text{max}}} 
%\approx \frac{2C_e}{\sqrt{\pi}}\frac{\Delta_0(\Delta_0-\Delta_1)R}{e f_a \mu_0 B_2 \epsilon_{z} D_S^P } \sqrt{\mathcal{P}Nt}\,.
%\end{aligned}
%\end{equation}
In the idealized experiment considered so far, since the axion signal $\eta$ is inversely proportional to the magnetic field $B_2$ whereas the SNR is independent of this quantity, one may suggest using arbitrarily small $B_2$ to enhance the axion signal. However, Eq. \eqref{SNR1} applies only if the photon noise makes the dominant contribution to the total noise. In practice, when the photon amplitude (proportional to $B_2$) becomes too small, signal-independent backgrounds will become significant, degrading the SNR. This determines the minimum usable value of $B_2$. Note also that close to this regime, the relative value of the interference term compared to the leading term in the transition probability is maximal.

\subsection{Atomic transition - the \texorpdfstring{$M1$}~\ case}\label{CalculationM1}
In this section, we present the calculation of the photon-axion interference term in the case of an $M1$ atomic transition.
\subsubsection{Photon absorption amplitude}
The $M1$ transition amplitude due to absorption of a photon is given by Eq.\ \eqref{General photon amplitude} but with the $M1$ photon matrix element given by
\begin{equation}\label{M1 photon matrix element}
M^{\gamma}_{BA}=\frac{e}{2m_e} \mathbf{\hat b}_{\gamma}\cdot\bra{B}\mathbf{J+S}\ket{A}%=\frac{\mathbf{\hat b}_{\gamma}\cdot\bra{B}\mathbf{S}\ket{A}}{2m_e}
\,,
\end{equation}
instead of Eq.\ \eqref{Photon matrix element}. Here, $\mathbf{\hat b}_{\gamma} = \mathbf{\hat k}_{\gamma} \times \boldsymbol{\epsilon}$, where $\mathbf{\hat k}_{\gamma}$ is the photon propagation unit vector (so $\mathbf{\hat k}_{\gamma}=\mathbf{\hat k}_a$), is the direction of the magnetic component of the photon field (mentioned above), $\mathbf{J}$ is the electron's total angular momentum and $\mathbf{S}$ is the electron's spin. %The second equality in Eq. \eqref{M1 photon matrix element} holds since the states $\ket{A}$ and $\ket{B}$ have different total angular momenta so $\bra{B}\mathbf{J}\ket{A}=0$.

If we now fix a spherical basis $\left\{\mathbf{e}_{-1},\mathbf{e}_{0},\mathbf{e}_{1}\right\}$ with the quantization axis $\mathbf{e}_0$ in the direction of the applied field $\mathbf{B}_2$, we can write the components of the vector $\mathbf{b}_{\gamma}$ as $b_{\gamma}^q$ where $q=-1,0,1$. We can also describe the states $A$ and $B$ by the  quantum numbers $n,j,l,m$ and $n',j',l',m'$, respectively. The $M1$ photon matrix element is then 
\begin{equation}\label{Photon M1 amplitude}
\begin{aligned}
M^{\gamma}_{BA}
=\left(-1\right)^{j'-m'}{\hat b}_{\gamma}^q\left(\begin{array}{ccc} j' & 1 & j\\
-m' & q & m\end{array} \right)\textfrak{J}\,,
\end{aligned}
\end{equation}
where $\left(\begin{array}{ccc} j' & 1 & j\\
-m' & q & m\end{array} \right)$ is the $3j$ symbol and $\textfrak{J}=\frac{e}{2m_e} \left<n'j'l'\|\mathbf{J+S}\|njl\right>$ is the reduced $M1$ matrix element. Note that here and below, summations over the repeated indices $p, q,...$ are implicit.

\subsubsection{Axion absorption amplitude}

The $M1$ transition amplitude due to absorption of an axion is given by Eq.\ \eqref{M-axion}, with the axion $M1$ matrix element given by
\begin{equation}\label{M1 axion amplitude}
\begin{aligned}
M^a_{BA}%=\frac{iC_e}{2f_a}\bra{B}\mathbf{\hat k}_a\cdot \boldsymbol{\sigma}\ket{A} \\
&= \frac{ig_{aee}}{m_e}\bra{B}\mathbf{\hat k}_a\cdot \mathbf{S}\ket{A}\\
&=\left(-1\right)^{j'-m'}\frac{ig_{aee}}{m_e}{\hat k}_{a}^q\left(\begin{array}{ccc} j' & 1 & j\\
-m' & q & m\end{array} \right)\textfrak{S}\,,
\end{aligned}
\end{equation}
where $\textfrak{S}= \left<n'j'l'\|\mathbf{S}\|njl\right>$ is the reduced matrix element of the operator $\mathbf{S}$. The second line of Eq. \eqref{M1 axion amplitude} is obtained by assuming the same spherical basis as above.
\subsubsection{Axion signal and signal-to-noise ratio}
Firstly, we prove that in the absence of the external field $\mathbf{B}_2$, the axion-photon interference term in the total transition probability vanishes. According to Eq.\ \eqref{P1}, this term is (twice) the product of the amplitudes \eqref{M1 photon matrix element} and \eqref{M1 axion amplitude}. When averaged over the initial projection $m$ and summed over the final projection $m'$, this term gives (up to a numerical factor)
\begin{equation}
\begin{aligned}
&\sum {\hat b}_{\gamma}^p {\hat k}_{\gamma}^q\left(\begin{array}{ccc} j' & 1 & j\\
-m' & p & m\end{array} \right)\left(\begin{array}{ccc} j' & 1 & j\\
-m' & q & m\end{array} \right)\\
&\propto \mathbf{\hat b}_{\gamma} \cdot \mathbf{\hat k}_{\gamma} \,,
\end{aligned}
\end{equation}
where the sum is over $m$, $m'$, $p$ and $q$. This quantity vanishes since $\mathbf{\hat b}_{\gamma}$ is perpendicular to $\mathbf{\hat k}_{\gamma}$ by construction. %This result can be understood if one note that, due to rotational symmetry, the axion matrix element must be a scalar, built from the relevant quantities, which, in this case, are the axion's momentum and the electron's spin; the only possibility is $\mathbf{b} \cdot \mathbf{n}_{\gamma}$(the other possibility ).

For the interference term to not average to zero, one may apply an external magnetic field $\mathbf{B}_2=B_2\mathbf{\hat z}$ to split the sublevels with different projections and tune the laser to induce transitions between levels of specific projections $\ket{j,m} \rightarrow \ket{j',m'}$. One can then define the axion signal $\eta$ as in Eq. \eqref{Signal 1}, but with photon and axion amplitudes defined by Eqs. \eqref{Photon M1 amplitude} and \eqref{M1 axion amplitude}. One gets 
\begin{equation}\label{M1 signal}
\begin{aligned}
\eta_{M1} &= \left| \frac{4\operatorname{Re}\left(i \overline{{{M}^{a}}}{{M}^{\gamma }} \right)}{\overline{{{M}^{\gamma }}}{{M}^{\gamma }}} \right|\\
&=\frac{2{g_{aee}}}{\sqrt{\pi }m_e}\left|\frac{{{\hat k}^{p}}\left( \begin{matrix}
   {{j}'} & 1 & j  \\
   -{m}' & p & m  \\
\end{matrix} \right)\textfrak{S}}{{{\hat b}^{q}}\left( \begin{matrix}
   {{j}'} & 1 & j  \\
   -{m}' & q & m  \\
\end{matrix} \right)\textfrak{J}}\right|\sqrt{\frac{\mathcal{P}}{\mathcal{T}}} \,.
\end{aligned}
\end{equation}

To calculate the SNR, one needs the (photon) absorption length of the target, which is given by Eq. \eqref{absorption length} but with $\Gamma_i=\frac{4}{3}\omega^3\left|M^{\gamma}_{BA}\right|^2$ being the rate of the $M1$ transition $\ket{j,m}\rightarrow \ket{j',m'}$ and $\Delta_0$ replaced by $\omega=E_{j',m'}-E_{j,m}$. The SNR is thus given by
\begin{equation}\label{M1 SNR}
\begin{aligned}
&\text{SNR}_{M1}=\frac{4g_{aee}}{m_e}\left|\frac{{{\hat k}_{\gamma}^{p}}\left( \begin{matrix}
   {{j}'} & 1 & j  \\
   -{m}' & p & m  \\
\end{matrix} \right)\textfrak{S}}{{{\hat b}_{\gamma}^{q}}\left( \begin{matrix}
   {{j}'} & 1 & j  \\
   -{m}' & q & m  \\
\end{matrix} \right)\textfrak{J}}\right|\sqrt{\frac{\mathcal{P}Nltn \Gamma_i}{ \omega^2\Gamma_{\rm tot}}} \\
&= \frac{4\sqrt{2}\pi^{1/4}{g_{aee}}}{\sqrt{3}{m_e}}\left|{{\hat k}_{\gamma}^{p}}\left( \begin{matrix}
   {{j}'} & 1 & j  \\
   -{m}' & p & m  \\
\end{matrix} \right)\textfrak{S}\right|\sqrt{\frac{\mathcal{P}Nltn}{v_0}}\,.
\end{aligned}
\end{equation} 
%where the parameters $P$, $N$, $l$, $t$, $n$ and $\Gamma_{\rm tot}$ are as in Sect.\ \ref{}. 
Here we assumed that the target atoms are in vapor form so that $\Gamma_{\rm tot}$ is given by the upper line in Eq.\ \eqref{Widths}.

We observe that just like in the $M0$ case, the SNR is independent of the wall's transmission coefficient $\mathcal{T}$.

%and the maximal SNR, achieved when $l \sim l_{\rm abs}$ so that all photons are absorbed, is given by
%\begin{equation}\label{M1 SNRmax}
%\begin{aligned}
%{\rm SNR_{max}}=\eta \sqrt{TN}\\
%=\frac{4{{C}_{e}}{{m}_{e}}}{\sqrt{\pi }e{{f}_{a}}}\left|\frac{{{\hat k}_{\gamma}^{p}}\left( \begin{matrix}
   %{{j}'} & 1 & j  \\
   %-{m}' & p & m  \\
%\end{matrix} \right)}{{{\hat b}_{\gamma}^{q}}\left( \begin{matrix}
   %{{j}'} & 1 & j  \\
   %-{m}' & q & m  \\
%\end{matrix} \right)}\right|\sqrt{PNt}\,.
%\end{aligned}
%\end{equation}
%We observe that the axion signal \eqref{M1 signal} and the maximal SNR \eqref{M1 SNRmax} do not depend on the reduced matrix element $\textfrak{S}_A^B$ and thus are independent of the type of the target atoms. The SNR \eqref{M1 SNR}, on the other hand, depends on the nature of the atoms via the energy difference $\omega$ and the width $\Gamma_i$ of the state $\ket{j',m'}$.
\subsection{Comparison of the two transition types}
In this section, we provide a rough comparison between the two mentioned transition types.

In the first scheme, where the axion-induced transition is of $M0$ type, the photon-induced transition amplitude is forbidden due to the selection rule $J=0 \nrightarrow J'=0$. In the second scheme, the photon-induced (as well as the axion-induced) amplitude is of $M1$ type and hence allowed. 

On the other hand, by inspection of Eq.\ \eqref{M_BA_exact} for the axion-induced transition matrix element, one deduces
\begin{equation}
\begin{aligned}
\left|\frac{M_{M0}^a}{M_{M1}^a}\right| &= \left|\frac{\omega_a\bra{B}\mathbf{r}\cdot \boldsymbol{\sigma }- \left({\mathbf{\hat k}}_{a}\cdot \boldsymbol{\sigma } \right) \left( {\mathbf{\hat k}}_{a}\cdot \mathbf{r} \right)\ket{A}}{\bra{B}\mathbf{\hat k}_a\cdot \boldsymbol{\sigma}\ket{A}}\right|\\
&\sim \omega_a r \sim  10^{-4} \mbox{--} 10^{-3} \,,
%r \ll 1
%\lesssim \alpha \,,
\end{aligned}
\end{equation}
where $\omega_a$ is the energy of the transition ($\lesssim \SI{}{\electronvolt}$), $r$ is a typical atomic radius ($\sim \SI{0.5}{\angstrom}$). Thus, the axion-induced $M0$ transition amplitude is much suppressed compared to its $M1$ counterpart.

These results mean that for some fixed set of experimental parameters, an axion-photon interference experiment which uses $M1$ transitions will generally give a larger absolute signal (which is proportional to the interference term in the total transition probability, i.e., the real part of the product of the axion- and photon-induced amplitudes) and thus might make it easier to detect axions than an experiment that uses $M0$ transitions. %However, as is shown in Sect. \ref{Discussion}, the relative axion signal (defined as the ratio between the axion- and photon-induced amplitudes) and the signal-to-noise ratio (SNR) of an interference experiment using $M1$ transitions are not in general better than those in an experiment that uses $M0$ transitions. 
In what follows, we provide calculations for both cases.
\section{Numerical estimates}\label{Discussion}
\subsection{Estimate of the axion signal and SNR - the \texorpdfstring{$M0$}~\ case}\label{DiscussionA}
We have derived the formulae for the signal and the SNR of the axion-photon $M0$ interference experiment, Eqs. \eqref{Signal 1} and \eqref{SNR2}, respectively. We now provide numerical estimates for these quantities. 

Besides the aforementioned situations where the target atoms are metals, we also consider the noble gases. Although the electronic configurations in the latter are different from those in the former (the outermost shell is $p^6$ instead of $s^2$), the calculation presented above is still good for the purpose of an estimate. We give numerical estimates for the noble gases using the same equations \eqref{Signal 1}, \eqref{absorption length} and \eqref{SNR2} as for metals.

For $\mathcal{P}$ and $N$, we use the values that are expected at the ALPS II experiment \cite{1748-0221-8-09-T09001}: $\mathcal{P} \sim 10^{-7}$ and $N \approx \SI{e20}{\per\second}$. We assume that $C_e\sim 1$ and $f_a = \SI{e9}{\giga\electronvolt}$, which corresponds to $g_{aee}=\SI{5e-13}{}$. For the transmission coefficient $\mathcal{T}$, we take $\mathcal{T}=10^{-18}$ to have the photon signal sufficiently suppressed. For simplicity, we assume that the laser light is polarized in the $z$-direction. 

The appropriate values for the atom density $n$ and temperature $T$ for different atoms are presented in Table \ref{Table 1} (see the caption of this table for more comments on the values of $n$ and $T$). We take $l=\SI{100}{\meter}$. %(the effective value of $l$ can be enlarged by using the cavity technique to trap the photons inside the target; axions are, of course, not affected)
For numerical estimates, we choose $t$=100 days, which is the order of magnitude for the maximal practical integration times. For $B_2$, we assume the value $\SI{e-4}{\tesla}$, which is slightly larger than Earth's magnetic field so no elaborate shielding is needed. The values for the reduced electric dipole matrix elements are given in \citep{PhysRevA.64.012508,NIST_ASD}. The resulting quantities are summarized in Table \ref{Table 1}.
\begin{widetext}
\begin{center}
\small
\centering
\begin{table}[htb]
\begin{tabular}{ |c|c|c|c|c|c|c|c|c|c|}
\hline
    Atom & $\Delta_0$ $\left(\SI{}{\electronvolt}\right)$ & $\Delta_1$ $\left(\SI{}{\electronvolt}\right)$ & $D_S^P/e$ $\left(\text{a.u.}\right)$ & $\tilde{D}_S^P/e$ $\left(\text{a.u.}\right)$ & $R$ $\left(\text{a.u.}\right)$ & $T$ $\left(\SI{}{\kelvin}\right)$ & $n $ $\left(\SI{}{\per\cubic\centi\meter}\right)$& $\eta_{M0}$ $\left(\times \SI{e-3}{}\right)$ & ${\rm SNR}_{M0}$ %& $l_a$ $\left(\SI{}{\kilo\meter}\right)$ 
    \\ \hline
    %Mg & 2.709 & 2.712 & 0.006 & 4.103 & $2.907$ & 1300 & $\SI{4.6e18}{}$ (v) & $\SI{3}{}$ & $\SI{0.07}{}$ %& $\SI{5e7}{}$
    %\\ \hline
    Ca & 1.879 & 1.886 & 0.03 & 4.93 & $3.52$ & 1700 & $\SI{4.3e18}{}$ (dv) &  $\SI{0.9}{}$ & $\SI{0.06}{}$ %& $\SI{4e7}{}$
    \\ \hline
    Sr & 1.775 & 1.798 & 0.15 & 5.39 & $3.96$ & 1600 & $\SI{5.1e18}{}$ (dv) & $\SI{0.8}{}$ & $\SI{0.1}{}$ %& $\SI{2e6}{}$
    \\ \hline
    Ba & 1.521 & 1.567 & 0.31 & 5.46 & $4.17$ & 2000 & $\SI{3.7e18}{}$ (dv) & $\SI{0.7}{}$ & $\SI{0.07}{}$ %& $\SI{4e7}{}$
    \\ \hline
    Hg & 4.667 & 4.886 & 0.45 & 2.64 & $2.32$ & 1000 & $\SI{2.2e19}{}$ (dv) & $\SI{4}{}$ & $\SI{0.4}{}$ %& $\SI{8e3}{}$ 
    \\ \hline
    Yb & 2.143 & 2.231 & 0.54 & 4.24 & $3.54$ & 1400 & $\SI{4.2e18}{}$ (dv) & $\SI{0.8}{}$ & $\SI{0.1}{}$ %& $\SI{4e5}{}$
    \\ \hline
    Ne & 16.72 & 16.85 & 0.60 & 0.17 & $0.73$ & 26 & $\SI{3.6e22}{}$ (lq) & $\SI{1}{}$ & $\SI{1}{}$ %& $\SI{30}{}$
    \\ \hline
    Ar & 11.72 & 11.83 & 0.93 & 0.46 & $1.14$ & 86 & $\SI{2.1e22}{}$ (lq) & $\SI{1}{}$ & $\SI{2}{}$ %& $\SI{70}{}$ 
    \\ \hline
    Kr & 10.56 & 10.64 & 0.85 & 0.91 & $1.04$ & 118 & $\SI{1.7e22}{}$ (lq) & $\SI{0.7}{}$ & $\SI{2}{}$ %& $\SI{10}{}$ 
    \\ \hline
        Xe & 9.447 & 9.570 & 0.89 & 1.15 & $1.09$ & 164 & $\SI{1.3e22}{}$ (lq) & $\SI{1}{}$ & $\SI{1}{}$ %& $\SI{40}{}$ 
        \\ \hline
    \end{tabular}
    \caption{Estimates of the $M0$ interference signal $\eta$ and SNR for some target atoms. For the metals, $\Delta_0$ and $\Delta_1$ are the energies of the states $nsnp\,{}^3P_0$ and $nsnp\,{}^3P_1$ with respect to the ground state $ns^2\,{}^1S_0$. For the noble gases, $\Delta_0$ and $\Delta_1$ are the energies of the states $np^5\,{}^2P_{1/2}\left(n+1\right)s\left[1/2\right]_0$ and $np^5\,{}^2P_{1/2}\left(n+1\right)s\left[1/2\right]_1$ with respect to the ground state $np^6\,{}^1S_0$. For the metals, $D_S^P$ and $\tilde{D}_S^P$ are the $E1$ reduced matrix elements between the ground state ${}^1S_0$ and the states ${}^3P_1$ and ${}^1P_1$, respectively. For the noble gases, $D_S^P$ and $\tilde{D}_S^P$ are the $E1$ reduced matrix elements between the ground state $np^6\,{}^1S_0$ and the states $np^5\,{}^2P_{1/2}\left(n+1\right)s\left[1/2\right]_1$ and $np^5\,{}^2P_{1/2}\left(n+1\right)s\left[3/2\right]_1$, respectively. We assume that the temperatures of the metals vapors (except for Hg) are only slightly lower than their corresponding boiling points. The densities $n$ of the vapors at these temperatures are estimated using the ideal gas equation and experimentally fitted vapor pressure equations presented in \cite{doi:10.1179/cmq.1984.23.3.309}. The temperature of the Hg vapor is taken to be $\SI{1000}{\kelvin}$ (higher than Hg's boiling point) and its density at this temperature is presented in \cite{doi:10.1063/1.433887}. The temperatures and densities of the noble gas liquids are presented in \cite{haynes2014crc}. The SNR corresponds to the target length of $\SI{100}{\meter}$ and integration time of $100$ days. The signal and SNR are presented for $g_{aee}=\SI{5e-13}{}$.}\label{Table 1}
\end{table}
\end{center}
\end{widetext}

We observe that for $g_{aee}=\SI{5e-13}{}$, the SNR in the cases of the nobles gases is of the order of unity. Thus, an axion-photon interference experiment that uses a noble gas as the axion and photon absorption medium is sensitive to the product $g_{a\gamma \gamma}g_{aee} \geq \SI{e-23}{\per\giga\electronvolt}$. If a metal vapor is used instead of a noble gas, the sensitivity decreases by two to three orders of magnitude.
%This is because the density of these elements (in their liquid form) is much higher than that of the metal vapors. Note also that it is technically easier to construct and maintain a long tube of liquid Xe at $\SI{164}{\kelvin}$ than that of Hg vapor at $\SI{1000}{\kelvin}$ for a long period of time. For example, at high temperature, the seals of the vapor cell deteriorates quickly so it is not possible to sustain the vapor's density in the long run. A tube of Xe at low temperature does not suffer from this issue. 
However, the drawback of the noble gases is their large excitation energies $\Delta_0$ and $\Delta_1$, which are far beyond the optical region. Such large energies can be achieved by using high-harmonic generation, but at the expense of the number of available photons.  %\textbf{(MORE COMMENT PLEASE!)}
%\subsection{Hyperfine structure}\label{Hyperfine}

%In the estimates presented in Table \ref{Table 1}, we assumed that the photon transition $\ket{^1S_0}\rightarrow \ket{^3P_0}$ is open due to the mixing of the states $\ket{^3P_0}$ and $\ket{^3P_1}$ in the magnetic field $B_2$, see eq.\ \eqref{admixture}. As mentioned earlier, the same mixing of states can be achieved with the hyperfine interaction, which becomes significant in isotopes with non-vanishing nuclear spin. It is interesting to compare these two effects. 

%Consider, for instance, the isotope $^{87}$Sr with the nuclear spin 9/2. Due to the hyperfine interaction, the natural width of the state $\ket{^3P_0}$ is $\Gamma_{^3P_0}\approx 0.0071\, {\rm s}^{-1}$ \cite{Zelevinsky}. The corresponding amplitude of the single-photon transition $\ket{^1S_0} \rightarrow \ket{^3P_0}$ reads $|M^\gamma|/\sqrt{n_1}\approx 2.16\times 10^{-9} {\rm eV}^{-1}$. Thus, the relative signal (\ref{Signal 1}) is now $\eta_{\text{Sr}}\approx 5\times 10^{-7}$. This value is approximately 150 times smaller than that given Table \ref{Table 1} for Sr because the photon amplitude is less suppressed. %However, we stress that because the hyperfine structure does not fluctuate, unlike the external magnetic field, it may give better precision in the measurement of the axion signal $\eta$.
\subsection{Estimate of the axion signal and SNR - the \texorpdfstring{$M1$}~\ case}\label{Discussion B}

We now provide numerical estimates for the axion signal and the SNR of the axion-photon $M1$ interference experiment, Eqs. \eqref{M1 signal} and \eqref{M1 SNR}. As explained in \cite{khriplovich1991parity}, the most suitable elements for an experiment involving atomic $M1$ transitions are Tl, Pb and Bi. We consider for Tl the transition $6s^26p\,{}^2P_{1/2} \rightarrow 6s^26p\,{}^2P_{3/2}$, for Pb the transition $6s^26p^2\,{}^3P_0 \rightarrow 6s^26p^2\,{}^3P_1$ and for Bi the transition $6s^26p^3\,{}^4S_{3/2} \rightarrow 6s^26p^3\,{}^2D_{3/2}$. % (these $M1$ transitions, along with others, were used to search for optical activity related to parity-nonconservation). %Note that since the total electronic angular momenta of the initial and final states in each of these transition are different, the reduced matrix element of the operator $\mathbf{J}$ vanishes so $\textfrak{J}=\frac{e}{2m_e}\textfrak{S}$ and the axion signal \eqref{M1 signal} is independent of the detail of the initial and final states. This means that the axion signal $\eta$ is the same for all three cases. 
The values of the reduced matrix element $\textfrak{J}$ for these transitions are presented in %${}^2P_{1/2} \rightarrow {}^2P_{3/2}$ in Tl can be calculated in an elementary way. The result is presented in \cite{khriplovich1991parity}. The values of $\textfrak{J}$ for the transition ${}^3P_{0} \rightarrow {}^3P_{1}$ in Pb and the transition ${}^4S_{3/2} \rightarrow {}^2D_{5/2}$ transition in Bi are presented in 
\cite{khriplovich1991parity,sushkov1979report,PhysRevA.44.2828}.

For simplicity of calculation, we consider for Tl the transition $\ket{j=1/2,m=1/2} \rightarrow \ket{j'=3/2,m'=1/2}$, for Pb the transition $\ket{j=0,m=0} \rightarrow \ket{j'=1,m'=0}$ and for Bi the transition $\ket{j=3/2,m=3/2} \rightarrow \ket{j'=3/2,m'=3/2}$. % (we assume that the magnetic field $\mathbf{B}_2$ has been applied to these atoms to lift the degeneracy of electronic levels with nonzero total angular momenta). %Note that a laser that induces the resonant transition $m=1/2 \rightarrow m'=1/2$ also induces the transition $m=-1/2 \rightarrow m'=-1/2$; however, since these two transitions cannot happen in the same atom, the axion signal is not affected; the total number of transitions is, on the other hand, doubled, so the SNR increases by a factor of $\sqrt{2}$. 
%In the case of Pb, when the field $\mathbf{B}_2$ is applied, the state with $J'=1$ splits into three states with $m'=0,\pm 1$. For simplicity of calculation, we assume that the laser is tuned to induce transitions between levels of projections $m=m'=0$. %Unlike in the case of Tl, there is no other transition with the same frequency so the SNR is not multiplied by any enhancement factor.
%In the case of Bi, when the field $\mathbf{B}_2$ is applied, the state ${}^4S_{3/2}$ splits into four states with $m'=\pm 3/2,\pm 1/2$ whereas the state ${}^2D_{5/2}$ splits into six states with $m'=\pm 5/2,\pm 3/2,\pm 1/2$. For simplicity of calculation, we assume that the laser is tuned to induce transitions between levels of projections $m=m'=1/2$. %There are three other transitions that have the same frequency, namely $m=-1/2 \rightarrow m'=-1/2$, $m=3/2 \rightarrow m'=3/2$ and $m=-3/2 \rightarrow m'=-3/2$. Thus, the SNR \eqref{M1 SNR} should be multiplied by a factor of $\sqrt{4}=2$.
For these transitions, we observe that the axion signal is proportional to the ratio $\left|{\hat k}_{\gamma}^0/{\hat b}_{\gamma}^0\right|$ so by arranging the photon's direction $\mathbf{\hat k}_{\gamma}$ very close to the $z$-axis (which is defined by the external field $\mathbf{B}_2$), one can make the axion signal large. For numerical estimates, we assume that ${\hat k}_{\gamma}^0 \approx 1$ and ${\hat b}_{\gamma}^0 \approx 0.01$.%, which corresponds to the angle $\left<\mathbf{\hat k}_{\gamma},\mathbf{B}_2\right> \approx 0.6^{\circ}$. 

For $g_{aee}$, $\mathcal{P}$, $N$, $\mathcal{T}$ and $t$, we assume the same values as in Sect. \ref{DiscussionA} whereas for the target length we can take $l=\SI{10}{\meter}$ (since the photon $M1$ transition is not forbidden, the photon absorption length will be small compared to the $M0$ case).
%With these values, the axion signal reads
%\begin{equation}
%\eta =\frac{4{{C}_{e}}{{m}_{e}}}{\sqrt{\pi }e{{f}_{a}}}\left|\frac{{{n}^{0}}}{{{b}^{0}}}\right|\sqrt{\frac{P}{T}} = \SI{4e-6}{} \left|\frac{{{n}^{0}}}{{{b}^{0}}}\right|\,.
%\end{equation}
%whereas the SNR reads
%\begin{equation}
%\text{SNR} = 
%\end{equation}
The results are summarized in Table. \ref{M1 results}.
%\begin{widetext}
\begin{center}
\small
\centering
\begin{table}[htb]
\begin{tabular}{ |c|>{\centering\arraybackslash}p{0.8cm}|>{\centering\arraybackslash}p{1.1cm}|>{\centering\arraybackslash}p{1.7cm}|>{\centering\arraybackslash}p{1.2cm}|c|}
\hline
    Atom & $\omega$ $\left(\SI{}{\electronvolt}\right)$ & $\textfrak{J}$ $\left(e/2m_e\right)$ %& $T$ $\left(\SI{}{\kelvin}\right)$ 
    & $n $ $\left(\SI{e17}{\per\cubic\centi\meter}\right)$& $\eta_{M1}$ $\left(\times\SI{e-4}{}\right)$ & ${\rm SNR}_{M1}$ %& $l_a$ $\left(\SI{}{\kilo\meter}\right)$ 
    \\ \hline
    Tl & 0.966 & -1.13 %& 1700 
    & $\SI{6.6}{}$  & $\SI{4.3}{}$ & $\SI{8.8}{}$ %& 9
    \\ \hline
    Pb & 0.969 & -1.29 %& 2000 
    & $\SI{1.1}{}$  &  $\SI{4.3}{}$ & $\SI{5.8}{}$ %& 200
    \\ \hline
    Bi & 1.416 & -1.69 %& 1800 
    & $\SI{1.5}{}$  & $\SI{4.3}{}$ & $\SI{6.0}{}$ %& 70
    \\ \hline
    \end{tabular}
    \caption{Estimates of the $M1$ interference signal $\eta$ and the SNR for some target atoms. Here, $\omega$ and $\textfrak{J}$ are the energy and $M1$ reduced matrix element of the transitions under consideration, respectively. The relevant values of $\textfrak{J}$ are given in \cite{khriplovich1991parity,sushkov1979report,PhysRevA.44.2828}. We assume that the temperature of the metal vapors is $\SI{1473}{\kelvin}$. The densities of the vapors at this temperature are estimated using the ideal gas equation and experimentally fitted vapor pressure equations presented in \cite{doi:10.1179/cmq.1984.23.3.309}. The SNR corresponds to the target's length of $\SI{10}{\meter}$ and integration time of $100$ days. The signal and SNR are presented for $g_{aee}=\SI{5e-13}{}$.}\label{M1 results}
\end{table}
\end{center}
%\end{widetext}
We observe that for $g_{aee}=\SI{5e-13}{}$, the SNR is significantly greater than unity. Thus, axion-photon interference experiments which use $M1$ transitions in post-transition metals are sensitive to the axion-electron coupling constant greater or of the order of $\SI{e-13}{} \mbox{--} \SI{e-12}{}$. %n fact, these experiments are also sensitive to the coupling constant down to $\SI{e-10}{\electronvolt}$. 
Overall, this scheme is sensitive to the product $g_{a\gamma \gamma }g_{aee}$ greater or of the order of $ \SI{e-24}{} \mbox{--} \SI{e-23}{\per\giga\electronvolt}$.
\subsection{Absorption and emission of axions by atoms}
In principle, the $M0$ transition $ {}^3P_0 \rightarrow {}^1S_0$ can be used to produce axions. Schematically, we can use a laser to resonantly excite the atoms in the ground state ${}^1S_0$ to some state with total angular momentum $J'=1$ then use another laser to bring the atoms in this state to the ${}^3P_0$ state. %Since the one-photon decay $\ket{{}^3P_0} \rightarrow \ket{{}^1S_0}$ is forbidden, we can keep the $\ket{{}^3P_0}$ state fully populated at all time. 
The state ${}^3P_0$ can only decay to the ground state ${}^1S_0$ by spontaneously emitting axions (or two photons). In this section, we give an estimate of the ${}^3P_0 \rightarrow {}^1S_0$ transition rate due to the spontaneous emission of axions. We also provide an estimate for the cross section of the absorption of an axion by a single atom (which makes the ${}^3P_0 \rightarrow {}^1S_0$ transition). This quantity might be of interest if one wishes to know how many axions are absorbed in the scheme proposed above.

The matrix element for the ${}^3P_0 \rightarrow {}^1S_0$ transition is obtained from Eq.\ \eqref{axion amplitude} by setting $n_a=1$. The transition rate then reads
\begin{equation}\label{axion emission rate}
\begin{aligned}
  \Gamma^a&=2\pi \int{\frac{\omega_1^2 d{{\omega }_{1} d \Omega }}{{{\left( 2\pi  \right)}^{3}}}\delta \left( {{\omega }_{1}}-{{\Delta }_{0}} \right){{\left| {{M}^{a}} \right|}^{2}}} \\ 
  &=\frac{\Delta _{0}^{2}}{{{\left( 2\pi  \right)}^{2}}}{{\left. {{\left| {{M}^{a}} \right|}^{2}} \right|}_{{{\omega }_{1}}={{\Delta }_{0}}}}=\frac{g_{aee}^2{{R}^{2}}\Delta _{0}^{5}}{9{{\pi }}m_e^{2}}\\
  &\approx {{\left( \frac{\SI{e9}{\giga\electronvolt}}{f_a/C_e} \right)}^{2}}{{\left( \frac{{{\Delta }_{0}}}{\SI{1}{\electronvolt}} \right)}^{5}} \left(\frac{R}{1 \text{ a.u.}}\right)^{2} \times \frac{3.86}{\SI{e30}{s}}   \,,
\end{aligned}
\end{equation}
where $\Delta_0$ is the energy of the ${}^3P_0$ state and the value of the integral $R$ is given in Table \ref{Table 1}. Here, we take $f_a/C_e = \SI{e9}{\giga\electronvolt}$%, which is the smallest permissible value consistent with astrophysical observations \cite{WhiteDwarf2008}
. The transition rates in some typical atoms are presented in Table \ref{transition rate due to axion emission in some atoms}.
%\begin{widetext}
\begin{center}
\begin{table}[htb]
\small
\centering
\begin{tabular}{|c|>{\centering\arraybackslash}p{1.8cm}|%>{\centering\arraybackslash}p{3cm}|
>{\centering\arraybackslash}p{1.8cm}|>{\centering\arraybackslash}p{1.7cm}|>{\centering\arraybackslash}p{1.7cm}| }
\hline
Atom & Axion spontaneous emission rate ${\Gamma^a}$ $\left( \SI{}{s^{-1}}\right)$ %& Natural width of the $\ket{{}^3P_0}$ state $\left( \SI{}{s^{-1}}\right)$ 
& Axion absorption cross section $\sigma_a$ $\left(a_B^2\right)$ &
Atom density ($\SI{}{\per\cubic\centi\meter}$) &
Number of axions produced ($\SI{}{\per\second\per\cubic\centi\meter}$)\\
\hline
Mg & $\SI{4.8e-27}{}$ %& $\SI{1.6e-13}{}$ 
& $\SI{7.7e-30}{}$ &$\SI{4.6e18}{}$ & $\SI{2.2e-8}{}$\\
\hline
Ca & $\SI{1.1e-27}{}$ %& $\SI{3.9e-13}{}$ 
& $\SI{6.3e-30}{}$ &$\SI{4.3e18}{}$ & $\SI{4.7e-9}{}$\\
\hline
Sr & $\SI{1.1e-27}{}$ %& $\SI{5.5e-12}{}$ 
& $\SI{1.5e-29}{}$ &$\SI{5.1e18}{}$ &  $\SI{5.6e-9}{}$\\
\hline
Ba & $\SI{5.5e-28}{}$ %&  
& $\SI{9.7e-30}{}$ &$\SI{3.7e18}{}$ &  $\SI{2.0e-9}{}$\\
\hline
Hg & $\SI{4.6e-26}{}$ %&  
& $\SI{4.8e-29}{}$ &$\SI{2.2e19}{}$ &  $\SI{1.0e-6}{}$ \\
\hline
Yb & $\SI{2.2e-27}{}$ %&  
& $\SI{1.9e-29}{}$ &$\SI{4.2e18}{}$ &  $\SI{9.2e-9}{}$\\
\hline
Ne & $\SI{2.6e-24}{}$ %& $\SI{2.3e-3}{}$ 
& $\SI{2.1e-30}{}$ &$\SI{3.6e22}{}$ &  $\SI{9.4e-4}{}$ \\
\hline
Ar & $\SI{1.1e-24}{}$ %& $\SI{2.2e-2}{}$ 
& $\SI{1.7e-30}{}$ &$\SI{2.1e22}{}$ &  $\SI{2.4e-4}{}$  \\
\hline
Kr & $\SI{5.5e-25}{}$ %& $2.0$ 
& $\SI{1.3e-30}{}$ &$\SI{1.7e22}{}$ &  $\SI{9.5e-5}{}$ \\
\hline
Xe & $\SI{3.5e-25}{}$ %& $12.8$ 
& $\SI{1.1e-30}{}$ &$\SI{1.3e22}{}$ &  $\SI{4.7e-5}{}$ \\
\hline
\end{tabular}
\caption{Estimates of the rates of the $ {}^3P_0 \rightarrow {}^1S_0$ transition due to spontaneous emission of axions, the axion absorption cross section and the number of axions emitted per unit time per unit volume. The axion absorption cross section is given in the units of $a_B^2$, where $a_B$ is the Bohr radius.} \label{transition rate due to axion emission in some atoms}
\end{table}
\end{center}
%\end{widetext}
Now suppose that one constructs an axion `laser' using one of these elements as the gain medium. Let us estimate the number of axions produced per second per unit volume ($\SI{}{\centi\meter}^3$) of the gain medium by this `laser'. This value is obtained by multiplying the transition rates (Table \ref{transition rate due to axion emission in some atoms}) by the number of atoms per unit volume ($\SI{}{\per\cubic\centi\meter}$) of the corresponding elements (Table \ref{Table 1}). The results are presented in the last column of Table \ref{transition rate due to axion emission in some atoms}. We observe that the number of axions produced (per unit time per unit volume) is small so an axion `laser' using the $J=0 \rightarrow J'=0$ is not efficient.

Finally, the axion absorption cross section $\sigma_a$ (without interference with a photon field) is obtained by replacing $\Gamma_i$ in Eq. \eqref{absorption length} with the axion emission rate $\Gamma^a$ \eqref{axion emission rate}. We find
\begin{equation}
\begin{aligned}
\sigma_a = \frac{4\pi}{\Delta_0^2} \frac{\Gamma^a}{\Gamma_{\text{tot}}}%= \frac{4g_{aee}^2R^2\Delta_0^3}{9m_e^2\Gamma_{\rm tot}} 
=\left\{\begin{array}{lcr}
\frac{2\sqrt{\pi}g_{aee}^2R^2\Delta_0^2}{9m_e^2v_0}  &\mbox{(dv)} \,,\\
\frac{2g_{aee}^2R^2\Delta_0^3}{9m_e^2v_0n\sigma_{\rm col}} &\mbox{(lq)} \,.
\end{array}\right. 
\end{aligned}
\end{equation}

%For a \textit{single, stationary} atom, the only contribution to $\Gamma_{\rm tot}$ is the natural width of $\ket{{}^3P_0}$, whose values for the atoms of interest are given in \cite{PhysRevA.69.042510,PhysRevA.11.1777}. Using these values, we obtain 
The estimate of this cross section $\sigma_a$ for some atoms are presented in Table \ref{transition rate due to axion emission in some atoms}.

%\subsection{Xe target}

%Above, we considered targets made of Mg, Ca and Sr atoms. It is possible to extend our calculation to the case if noble gases such as Xe. One advantage of having a Xe target is that since there are many intermediate levels lying between the ground state and the first excited state with total angular momentum zero, it is no longer necessary to use a second laser to further excite the target atoms after they absorb axions: these atoms will de-excite down to some lower-lying level and the resulting fluorescent can be detected. Another advantage that a Xe target has over Mg, Ca and Sr targets is the higher density.
\section{Conclusions}
In this work, we have proposed two schemes for resonant detection of laboratory-produced axions and other axion-like particles. In our schemes, the axions are generated from photons in a magnetic field, as in current LSW experiments, and are then detected by using atomic (or molecular) transitions. The fundamental difference between our schemes and traditional LSW experiment is that instead of completely blocking off the photons, we allow a fraction of them to be absorbed by the target atoms. With such an allowance, the interference between the axion- and photon-induced transition amplitudes occurs and the experimental signal now scales linearly with the axion-electron coupling constant. This is an improvement over existing atom-based proposals whose signals have a quadratic dependence on the axion-electron coupling constant.

\begin{figure}[h]
\includegraphics[scale=0.42]{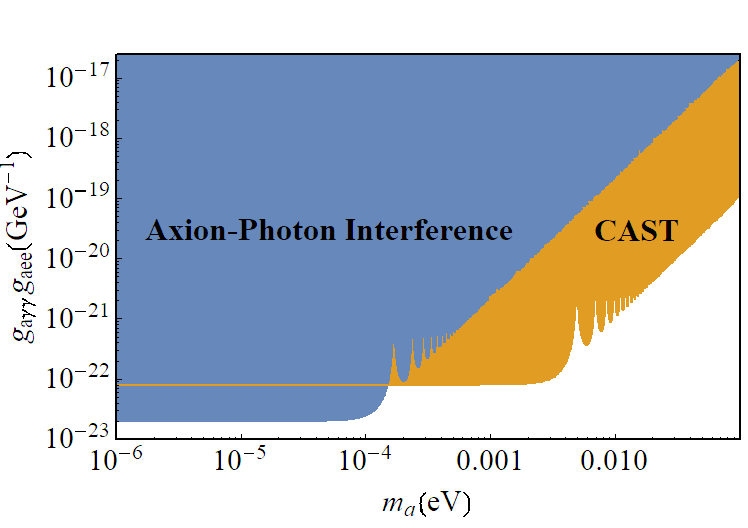}
\caption{Comparison between constraints on the product $g_{a\gamma \gamma}g_{aee}$ as a function of axion mass $m_a$ as fixed by CAST \cite {1475-7516-2013-05-010} and by the experimental scheme proposed in this paper. For small axion mass $m_a \lesssim \SI{e-4}{\electronvolt}$, our projected constraint is better whereas for large axion mass $m_a > \SI{e-4}{\electronvolt}$, CAST seems to be more competitive.}\label{CAST Interference graph}
\end{figure}

We have provided theoretical calculations and numerical estimates for a number of target atoms. We found that noble gases, in which axions induce transitions of $M0$ type, and post-transition metals, in which axions induce transitions of $M1$ type, are potential candidates for experimental applications. These schemes may be realized as simple upgrades of the existing and planned ALPS experiments (the photon-blocking wall replaced by some semi-transparent material and the axion-to-photon reconversion unit replaced by a vapor cell). The proposed schemes have a sensitivity to the product of the axion-photon and axion-electron coupling constants $g_{a\gamma \gamma}g_{aee}$ greater or of the order of $\SI{e-24}{} \mbox{--} \SI{e-23}{\per\giga\electronvolt}$. A comparison between this and the value constrained by CAST observational data \cite {1475-7516-2013-05-010} is presented in Fig.\ \eqref{CAST Interference graph}. As can be seen, for small axion mass $m_a \lesssim \SI{e-4}{\electronvolt}$, our limit will be more stringer than the limit set by CAST. However, for larger mass $m_a > \SI{e-4}{\electronvolt}$, CAST seems to be more sensitive. It is also of interest to mention that the gaps in the sensitivity due to `wiggles' at large axion mass can be partially eliminated by using buffer gas to produce a photon refractive index $n>1$ \cite{PhysRevD.39.2089,EHRET2010149}.

We note that the sensitivity to $g_{a\gamma \gamma}g_{aee}$ might be enhanced in an upgraded version of our experiment which employs heterodyne interferometry, whose usefulness to ALSP-type experiments was studied in \cite{Bush:2017yuk}. The authors of \cite{Bush:2017yuk} suggested to interfere the photon signal in an ALPS-type experiment with some laser of slightly different frequency. This generates a time varying signal, called the beat note (at the difference frequency) which is separated from the constant photon background and is detected. The authors hoped to achieve a sensitivity of $10^{-6}$ photon per second. Following this idea, we can consider a scheme in which the axion-generating photons are blocked off completely (a completely opaque wall is used instead of a semi-transparent one) and the axion-induced transition amplitude is allowed to interfere with the amplitude of the transition caused by photons of slightly different frequency. The beat note in the total number of transitions is detected. This scheme may have some advantages over those considered in this paper. A detail study of it will the the subject of our future work.

As a side result to this paper, we also studied the possibility of using atomic transitions to produce axions. We found that this type of axion production is not as effective as converting photons into axions in a magnetic field. A possibility of coherent enhancement of the photons production by axions and axions production by photons in the forward direction will be considered in a separate publication.

\section*{Acknowledgments}
We thank Carlo Rizzo for asking the right questions that triggered this work and for his advice on the manuscript and Vladimir Dzuba and Max Zolotorev for helpful discussions. This work was supported in part by the Australian Research Council, the Gutenberg Fellowship, the Humboldt Research Fellowship, the DFG Koselleck Program and the Heising-Simons and Simons Foundation. This project has also received funding from the European Research Council (ERC) under the European Union's Horizon 2020 Research and Innovative Programme (grant agreement No. 695405).

\bibliography{LitM1}
\end{document}